\documentstyle[aps,prd,floats,epsf]{revtex}
\begin{document}

\def\dt{{\hat \partial}_0}
\def\dr{{\partial\over\partial r}}

\def\gamr{{\Gamma^r{}_{rr}}}
\def\dgamr{{\partial{\Gamma^r{}_{rr}}\over\partial r}}
\def\grt{{\Gamma_{rT}}}
\def\mrr{{M_{rrr}}}
\def\mtr{{M_{Tr}}}
\def\mrt{{M_{rT}}}
\def\kr{{K_{rr}}}
\def\kt{{K_T}}
\def\lr{{L_{rr}}}
\def\lt{{L_T}}
\def\ar{{a_r}}
\def\arr{{a_{rr}}}
\def\at{{a_T}}
\def\aor{{a_{0r}}}
\def\krtil{{K^r{}_r}}
\def\drr{{\partial\tilde{r}\over\partial r}{\partial\over\partial\tilde r}}
\def\dtt{{\partial\over\partial\tilde{t}}}
\def\jacoby{{\partial\tilde{r}\over\partial r}}

\preprint{CRSR 1122}
\draft 

\title{Numerical Evolution of Black Holes
with a Hyperbolic Formulation of General Relativity}
\author{Mark A.\ Scheel$^1$, Thomas W.\ Baumgarte$^2$,
        Gregory B.\ Cook$^1$,\\
	Stuart L.\ Shapiro$^{2,3}$, and Saul A.\ Teukolsky$^{1,4}$}
\address{$^1$Center for Radiophysics and Space
Research,Cornell University, Ithaca, New York\ \ 14853}
\address{$^2$Department of Physics, University of
Illinois at Urbana-Champaign, Urbana, IL\ \ 61801}
\address{$^3$Department of Astronomy and
NCSA, University of Illinois at Urbana-Champaign, Urbana, IL\ \ 61801}
\address{$^4$Departments of
Physics and Astronomy, Cornell University, Ithaca, New York\ \ 14853}

\date{\today}

\maketitle

\begin{abstract}
\widetext
We describe a numerical code that solves Einstein's equations for a
Schwarzschild black hole in spherical symmetry, using a hyperbolic
formulation introduced by Choquet-Bruhat and York. This is the first
time this formulation has been used to evolve a numerical spacetime
containing a black hole. We excise the hole from the computational
grid in order to avoid the central singularity.  We describe in detail
a causal differencing method that should allow one to stably evolve a
hyperbolic system of equations in three spatial dimensions with an
arbitrary shift vector, to second-order accuracy in both space and
time. We demonstrate the success of this method in the spherically
symmetric case.
\end{abstract}

\section{Introduction} \label{sec:intro}

A key goal of numerical relativity is to determine the gravitational
radiation produced by the inspiral and coalescence of two black holes
in a decaying binary orbit.  The importance of solving this problem is
heightened by the possibility that gravitational wave detectors such
as LIGO, VIRGO, and GEO may observe gravitational waveforms from
binary black hole coalescence within the next decade.  Not only could
a comparison of measured waveforms with numerical simulations provide
a crucial strong-field test of general relativity, but in addition,
accurate templates produced by these simulations could significantly
increase the sensitivity of waveform measurements \cite{thorne96} and
reduce the uncertainties in astrophysical parameters derived from
gravitational wave data.

Although constructing a numerical simulation of binary black hole
coalescence is a difficult and unsolved problem, several recent
advances have brought us closer to a solution.  One key advance is the
development of so-called apparent horizon boundary condition (AHBC)
methods \cite{seidel_suen92,anninos_etal95,scheel95a,scheel95b,%
bona_masso95a,marsa96}, which treat black holes by evolving only the
regions of spacetime that lie outside apparent horizons.  These
methods take advantage of the fact that information cannot emerge from
within the apparent horizon of a black hole (which, assuming cosmic
censorship, is contained within the event horizon).  Without AHBC
schemes, the spacetime singularity that inevitably forms inside a
black hole eventually causes numerical simulations to terminate,
typically on a time scale of order $10$ to $100 M$, where $M$ is the
mass of the system.  For this reason, AHBC methods may be crucial for
solving the binary black hole problem, where one hopes to evolve two
holes long enough to see them orbit, coalesce, and eventually settle
into a final state containing a single (Kerr) black hole.

Another promising development is the construction of manifestly
hyperbolic formulations of Einstein's equations
\cite{frittelli_reula94,choquet_york95,abrahams_etal95,%
bona_masso95b,mvp96,friedrich96}.  Not only do these formulations
offer insights into the mathematical structure of general relativity,
but they may also be better-suited for numerical solution than the
usual ADM\cite{ADM} equations, which are not manifestly hyperbolic.
The reason for this is twofold: First, there exists an extensive
literature concerning stable and efficient numerical methods for
solving hyperbolic systems of equations\cite{leveque}. Some of these
methods have been successfully applied to hyperbolic formulations of
general relativity by Bona and Masso\cite{bona_masso95a}.  The second
reason is that a non-hyperbolic set of equations can present a
fundamental difficulty for black hole simulations that employ an AHBC
approach.

To understand this difficulty, consider a non-hyperbolic set of
equations, or even a hyperbolic set that has characteristics lying
outside the local light cone.  Suppose such a system is to be solved
on a domain that includes a black hole.  Although the physics
guarantees that nothing can emerge from the hole, the equations do not
know this, and nonphysical (gauge) modes can propagate outward through
the apparent horizon. Solving such a system of equations on a
restricted domain that excludes the interior of the hole is
mathematically well-posed only if appropriate boundary conditions are
imposed on the horizon.  While it should be possible to impose
explicit horizon boundary conditions to fix the coordinate
system\cite{eardley97}, it is unclear which boundary conditions are
appropriate for dynamical variables, particularly in the general
three-dimensional case in which one may have a nonspherical horizon
and a significant amount of gravitational radiation.

Now consider a hyperbolic set of equations with characteristics that
never lie outside the local light cone.  In this case, future-pointing
characteristics inside the apparent horizon (which must be an outgoing
non-timelike surface) can never intersect the horizon itself, so that
quantities at or outside the horizon cannot depend on the interior
region.  Consequently, one can solve this set of equations on a
restricted domain that excludes the interior of the hole, and one can
do so without imposing boundary conditions on the horizon.

In this paper, we concentrate on the hyperbolic formulation of
Einstein's equations originally proposed by Choquet-Bruhat and York
\cite{choquet_york95,abrahams_etal95}, hereafter referred to as the
CBY system.  This formulation has several advantageous features.
First, the characteristics of this set of equations are extremely
simple: they lie either along the light cone or along the normal to
the current time slice. This guarantees that no information, not even
gauge information, can propagate acausally. Second, the equations
admit an arbitrary shift vector, and the characteristics are
independent of the choice of shift. Finally, the fundamental variables
in this formulation are spatially covariant three-dimensional
tensors. These tensors directly measure spacetime curvature, and from
them one can form all components of the spacetime Weyl tensor.

While we believe that the CBY formalism holds considerable promise for
numerical simulations, particularly when combined with AHBC methods,
there is no experience with solving this particular set of equations
on a computer.  Therefore, before expending the significant effort
required to implement the CBY equations in a full three-dimensional
code, it is important to demonstrate that such an approach is
feasible.

Accordingly, we have developed a numerical code that solves the
spherically symmetric Einstein equations using the CBY formalism.  We
evolve a Schwarzschild spacetime using a causal differencing scheme,
and we use an AHBC method to avoid the central singularity.  This is
the first time the CBY equations have been used to evolve a numerical
spacetime containing a black hole.

Our code runs in parallel, and is rigorously second order convergent.
As described in detail elsewhere\cite{scheel_etal97b}, it can run for
times in excess of $1000M$ provided certain constraints are regularly
enforced.  Our code is based on the DAGH\cite{dagh} software package
originally developed for the Binary Black Hole Grand Challenge
Alliance. It is written in C++, and uses Fortran-90 numerical kernels.
The DAGH system contains support for adaptive mesh refinement, but we
have not yet taken advantage of this feature.

Our code provides an important demonstration that the CBY formulation
works well in numerical simulations.  It allows us to study the
details of implementing the CBY equations in a simple setting, and it
provides a testing ground for apparent horizon excising schemes and
causal differencing algorithms.  It also serves as an important check
on a code that solves the CBY equations in three spatial dimensions, a
code that is currently under development and will be described
elsewhere.

We employ results and methods specific to spherical symmetry as little
as possible, so that our techniques are readily generalizable to the
three-dimensional case with Cartesian coordinates. For example, we do
not use logarithmic radial coordinates\cite{seidel_suen92,scheel95a}
despite the great advantages they provide for spherically symmetric
codes.  Likewise, although there are many shift conditions that work
well with AHBC methods in spherical symmetry\cite{anninos_etal95}, we
consider only those that can be applied in the general
three-dimensional case.

Recent progress has also been made in spherical symmetry by choosing
a global coordinate system that has desirable properties near the
horizon\cite{marsa96}.  The spatial gauge used by \cite{marsa96}
depends on the concept of an areal radius, and is thus applicable only
in spherical symmetry.  We forego such a coordinate choice in favor of
a more general approach.

In section~\ref{sec:eqns}, we summarize the CBY formulation of
Einstein's equations, and we specialize this formulation to the case
of spherical symmetry.  In section~\ref{sec:evolution}, we present a
key ingredient of our code: a causal differencing scheme that is
second-order convergent and has a stability criterion that is
independent of the shift vector.  We describe this scheme in detail
for the general case of three spatial dimensions, and then apply it to
the spherically symmetric case.  In section~\ref{sec:shift} we
describe the AHBC method employed at the inner boundary of our grid,
and the shift conditions that we use in order to implement this
method.  These shift conditions not only allow us to control the
motion of the apparent horizon through the grid, but they also prevent
coordinate singularities that may result from differential stretching
or compression of the grid in the remainder of the spacetime.  In
section~\ref{sec:outerbdry} we discuss the boundary condition that we
impose at the outer boundary of our grid.  This boundary is ideally at
spatial infinity, but in practice it is placed at a large but finite
radius.  In section~\ref{sec:tests} we present the results of rigorous
convergence tests using our code.  In section~\ref{sec:conclusion}, we
close with a short discussion of our results.

\section{Equations}\label{sec:eqns}
\subsection{The CBY Formalism} \label{sec:eqns.ny}
Here we summarize the fundamental equations and variables used in the
CBY representation of general relativity.  For details of the CBY
formulation and a derivation of the equations, see
\cite{choquet_york95}.

We write the metric in the usual 3+1 form
\begin{equation} \label{fourmetric}
ds^2=-N^2dt^2+g_{ij}(dx^i+\beta^idt)(dx^j+\beta^jdt),
\end{equation}
where $N$ is the lapse function, $\beta^i$ is the shift vector, and
$g_{ij}$ is the three-metric on a spatial hypersurface of constant
$t$.

Define the variables
\begin{mathletters}\label{yorkvariables}
\begin{eqnarray}
K_{ij} &\equiv& -{1\over 2}N^{-1}\dt g_{ij},\\
L_{ij} &\equiv& N^{-1}\dt K_{ij},\\
M_{kij}&\equiv& D_kK_{ij},\label{def:mkij}\\
a_i    &\equiv& D_i(\ln N),\label{def:ai}\\
a_{0i} &\equiv& N^{-1}\dt a_i,\\
a_{ij} &\equiv& D_ja_i\label{def:aij}.
\end{eqnarray}
\end{mathletters}
Here $D$ is the three-dimensional covariant derivative compatible with
the three-metric $g_{ij}$, the time derivative operator is
\begin{equation} \label{dt}
\dt\equiv {\partial\over\partial t}-\pounds_\beta,
\end{equation}
and \pounds\ denotes a Lie derivative. The quantity $K_{ij}$ is the
usual extrinsic curvature.

If we assume that the time coordinate satisfies the harmonic slicing
condition
\begin{equation}
     \Box t = 0, \label{eqn:harmonicslicing}
\end{equation}
then the vacuum evolution equations take the form
\begin{mathletters} \label{genevolution}
\begin{eqnarray}
\dt g_{ij} &=& -2N K_{ij} \label{gijevolution}
\\
\dt K_{ij} &=& N L_{ij} \\
\dt a_i &=& N a_{0i} \\
\dt N &=& -N^2 H \label{harmonic_slicing}\\
\dt {\Gamma^k}_{ij} &=& -N(2a_{(i} K_{j)}^k - a^k K_{ij} +
        2 {M_{(ij)}}^k - {M^k}_{ij}) \label{gammaevolution}
\end{eqnarray}
\begin{eqnarray}
\dt L_{ij} - N D_k{M^k}_{ij} &=& -NJ_{ij} \label{lijevolution}\\
\dt M_{kij} - N D_k L_{ij} &=& N\Bigl[2\Bigl(
        a_k K^\ell_{(i} K_{j)\ell} + K_{k\ell} K^\ell_{(i} a_{j)}
        - K_{k(i} K_{j)}^\ell a_\ell 
\nonumber\\ &&\mbox{}
        + K^\ell_{(i} M_{j)k\ell} 
        + M_{k\ell(i} K_{j)}^\ell - M_{\ell k(i} K_{j)}^\ell\Bigr)
        + a_k L_{ij} \Bigr] \\  
\dt a_{0i} - N D_j a_i^j &=& NQ_i\\
\dt a_{ij} - N D_j a_{0i} &=& N\left[
        a_j a_{0i} + 2a_{(i} K_{j)}^{k}a_k + 2{M_{(ij)}}^k a_k
\right. \nonumber\\ && \left.\mbox{}
        - a^k a_k K_{ij} - a_k{M^k}_{ij}\right] \label{aijevolution}
\end{eqnarray}
where $H$ is the trace of $K_{ij}$, $X_{(ij)}\equiv(X_{ij} +
X_{ji})/2$, the quantities ${\Gamma^k}_{ij}$ are the connection
coefficients associated with the spatial covariant derivative $D_k$,
and
\begin{eqnarray}
J_{ij} &\equiv& g_{ij}\left[H(L - H^2 + a^k a_k + a^k_k)
\right.\nonumber\\ && \left.\mbox{}
        + K^{k\ell}(4H K_{k\ell} -2L_{k\ell} - 4K_{mk} K^m_\ell
        - 2a_k a_\ell -2a_{k\ell})\right] \nonumber\\ &&\mbox{}
        - K_{ij}(L - 3H^2 + a^k_k + 2a^k a_k + 3K^{k\ell}K_{k\ell})
\nonumber\\ &&\mbox{}
        - H(3L_{ij} + 6a_i a_j + 4a_{ij} + 10K_{ik} K^k_j)
\nonumber\\ &&\mbox{}
        + 2K^k_{(i}\left[5L_{j)k} + 8K_{j)\ell} K_k^\ell
        + 5a_{j)} a_k + 4a_{j)k}\right] 
\nonumber\\ &&\mbox{}
        + a_k(4{M_{(ij)}}^k
        - 3{M^k}_{ij}) - 4a_{(j}{M_{i)k}}^k \\
Q_i &\equiv& H{M_{ij}}^j - 2K^{jk}M_{ijk} + a_j\left(
        a_i^j - L_i^j + H K_i^j - 2K_{ik}K^{kj}\right)
\nonumber\\ &&\mbox{}
        + a_i\left(H^2 + a^j a_j + 2a^j_j - 2K^{jk}K_{jk}\right).
\end{eqnarray}
\end{mathletters}
Eq.~(\ref{harmonic_slicing}) is the harmonic slicing
condition~(\ref{eqn:harmonicslicing}).

There are considerably more variables and equations in the CBY
formalism than in the usual ADM formalism. However, the form of the
equations is much simpler in the CBY case.  While the right-hand sides
of Eqs.~(\ref{genevolution}) contain many terms, these terms consist
solely of algebraic combinations of the dynamical variables and
involve no derivatives.  Eqs.~(\ref{lijevolution}--\ref{aijevolution})
are tensor wave equations whose characteristics are along the light
cones.  Eqs.~(\ref{gijevolution}--\ref{gammaevolution}) are even
simpler---they drag information normal to the surfaces of constant
$t$, that is, along zero-velocity (with respect to the normal)
characteristics.  There are no other characteristics in the system.

Although we have eliminated some gauge freedom in the choice of lapse
function by imposing the harmonic time slicing
condition~(\ref{eqn:harmonicslicing}), the shift $\beta^i$ is
unspecified and completely arbitrary.  The shift is not a dynamical
variable in this formalism, in the sense that it obeys no evolution
equation, and that it appears in Eqs.~(\ref{genevolution}) only
through the time derivative operator $\dt$.  Instead, the shift is an
auxiliary gauge variable that may be freely chosen on each time slice,
and may even change discontinuously from one slice to the next.

In vacuum, the constraint equations include
\begin{mathletters} \label{genconstraints}
\begin{eqnarray} 
0  &=& L_i^i + K^{ij}K_{ij} + a^i a_i + a_i^i \label{genHamConst}\\
0  &=& {M^j}_{ji} - {M_{ij}}^j \label{genMomConst} \\
0  &=& a_{0i} + H a_i + {M_{ij}}^j \label{genAoiDef}\\
0  &=& \bar{R}_{ij} - L_{ij} + H K_{ij} - 2K_{ik} K^k_j - a_i a_j - a_{ij} 
         \label{genkijadmevolution},
\end{eqnarray}
where $\bar{R}_{ij}$ is the three-dimensional Ricci tensor.
Eqs.~(\ref{genHamConst}) and~(\ref{genMomConst}) are the familiar
Hamiltonian and momentum constraints rewritten in terms of the CBY
variables, and Eq.~(\ref{genAoiDef}) is a result of harmonic time
slicing.  Eq.~(\ref{genkijadmevolution}) is the familiar ADM evolution
equation for $K_{ij}$, which in the CBY picture becomes a constraint
on $L_{ij}$.  Eqs.~(\ref{def:mkij}), (\ref{def:ai}), (\ref{def:aij}),
and the usual relation between $\Gamma^k{}_{ij}$ and derivatives of
$g_{ij}$ are also constraints that must be satisfied at all times.
All constraints are preserved by the evolution equations.
\end{mathletters}

\subsection{Spherical Symmetry} \label{sec:eqns.ss}
 
We write the spherically symmetric three-metric in the general form
\begin{equation} \label{3metric}
{}^{(3)}ds^2 = A^2dr^2+B^2r^2(d\theta^2+\sin^2\theta\,d\phi^2),
\end{equation}
where ($r,\theta,\phi$) are the usual spherical coordinates.
Spherical symmetry reduces the number of dynamical variables
(including the connection coefficients)
from 67 to 16.
Define
\begin{mathletters} \label{sphericalvariables}
\begin{eqnarray}
\grt   &\equiv& 2 Br\Gamma^\theta{}_{\theta r}
              = 2 Br\Gamma^\phi{}_{\phi r}     
              =-{2A^2\over Br}\Gamma^r{}_{\theta\theta}
              =-{2A^2\over Br \sin^2\theta}\Gamma^r{}_{\phi\phi}\\
\at    &\equiv& a^\theta{}_\theta = a^\phi{}_\phi\\
\lt    &\equiv& L^\theta{}_\theta = L^\phi{}_\phi\\
\kt    &\equiv& K^\theta{}_\theta = K^\phi{}_\phi\\
\mrt   &\equiv& M_r{}^\theta{}_\theta = M_r{}^\phi{}_\phi\\
\mtr   &\equiv& M^\theta{}_\theta{}_r = M^\phi{}_\phi{}_r,
\end{eqnarray}
\end{mathletters}
where the subscript $T$ denotes ``transverse''.  Then the vacuum
evolution equations~(\ref{genevolution}) take the form
\begin{mathletters} \label{CBYespher}
\begin{eqnarray}
\dt A    &=& -NA\krtil\label{aevolution}\\
\dt Br   &=& -NBr\kt\\
\dt \kr  &=& \phantom{-}N\lr\\
\dt \kt  &=& \phantom{-}N(\lt+2\kt^2)\\
\dt N    &=& -N^2(\krtil+2\kt)\label{nevolution}\\
\dt a_r  &=& \phantom{-}N\aor\\
\dt \at  &=& \phantom{-}N\left[(2\mtr-\mrt-\ar\kt){\ar\over A^2}
                    \nonumber \right. \\ &&\qquad\qquad\left.
              +{\grt\over 2A^2Br}\aor
              +2\kt\at\right]\\
\dt \gamr &=& -{N\over A^2}\left[\kr\ar+\mrr\right]\\
\dt \grt &=& -N\left[\kt\grt+2Br(\ar\kt+\mrt)\right]\\
\dt \mtr &=& \phantom{-}N\left[\kt(2\mtr+\mrt+\ar\kt)
              +{\grt\over 2Br}\left({\lr\over A^2}-\lt\right)
                     \nonumber \right. \\ &&\qquad\left.
              +\vphantom{\lr\over A^2}\krtil(2\mtr-\mrt-\ar\kt)\right]
              \label{mtrevolution}
\\
\dt \arr   -   N\dr\aor
                     &=&    N\left[-\gamr\aor
                           +\ar(\ar\krtil+\mrr/A^2+\aor)\right]
                           \label{arrevolution}\\
\dt \aor   -   {N \over A^2}\dr\arr
                     &=&    N\left[Q_r-{2\gamr\arr\over A^2}
                           +{\grt\over Br}({\arr/A^2}-\at)\right]\\
\dt \mrr   -   N\dr\lr 
                     &=&    N\left[(\ar-2\gamr)\lr
                           +2\krtil(\kr\ar+\mrr)\right]\\
\dt \mrt   -   N\dr\lt
                     &=&    N\left[2\kt(\ar\kt+2\mrt)+\ar\lt\right]\\
\dt \lr    -   {N \over A^2}\dr\mrr
                     &=&    N\left[-J_{rr}-{3\gamr\mrr\over A^2}
                           +{\grt\over Br}(\mrr/A^2-2\mtr)\right]\\
\dt \lt    -   {N \over A^2}\dr\mrt
                     &=&    N\left[-J_{T}
                           -{\gamr\mrt\over A^2}
                           +{\grt\over A^2 Br}(\mrt+\mtr)\right],
                           \label{ltevolution}
\end{eqnarray}
where
\begin{eqnarray}\label{right_hand_sides_begin}
J_{T}  &\equiv&  (\lr+\arr)(\kt-\krtil){1\over A^2} 
                 -\lt\krtil
                 -{\ar^2\krtil\over A^2}
                      \nonumber \\ &&\qquad
                 -2\at(\krtil+\kt)
                 +2\kt^3-2\krtil\kt^2
                      \nonumber \\ &&\qquad
                 +2\krtil^2\kt-\krtil^3
                 +{\ar\over A^2}(4\mtr-3\mrt),\\
J_{rr} &\equiv&  \lr(5\krtil-4\kt)
                 +\ar^2(\krtil-10\kt)
                 +2\arr(\krtil-3\kt)
                      \nonumber \\ &&\qquad
                 +\kr(5\krtil^2-6\krtil\kt+2\kt^2)
                 -\ar\left({3\mrr\over A^2}+8\mrt\right),\\
Q_r    &\equiv&  {\mrr\over A^2}(2\kt-\krtil) +2\mrt\krtil
                 +\ar\left((\ar^2+3\arr-\lr){1\over A^2}
                      \nonumber \right. \\&&\qquad\left.\vphantom{1\over A^2}
                 +4\at-2\krtil(\krtil-3\kt)\right).
                  \label{right_hand_sides_end}
\end{eqnarray}
\end{mathletters}
The constraints~(\ref{genconstraints}) become
\begin{mathletters}
\label{sphersymconstraints}
\begin{eqnarray}
     2\lt+{\lr\over A^2}+2\kt^2+\krtil^2+2\at
                        +{1\over A^2}(\ar^2+\arr) &=& 0
       \label{HamConst}
     \\ 
        \mrt - \mtr   &=& 0
        \label{MomConst}
     \\	
        \aor + \ar(2\kt+\krtil)+{\mrr\over A^2}+2\mrt &=& 0
	\label{AorDef}
     \\
        {1\over Br}\left[-\dr\grt + \gamr\grt\right]
        -\arr-\ar^2+\kr(2\kt-\krtil)-\lr &=& 0
        \label{krradmevolution}
     \\
      {1\over 2 A^2 Br}\left[-\dr\grt+\gamr\grt-{\grt^2\over 2 Br}\right]
	     +{1\over B^2r^2}+\kt\krtil-\at-\lt &=& 0.
        \label{ktadmevolution}
\end{eqnarray}
\end{mathletters}

The additional constraints~(\ref{def:mkij}), (\ref{def:ai}),
(\ref{def:aij}), and the usual relation between $\Gamma^k{}_{ij}$ and
derivatives of $g_{ij}$ take the form
\begin{mathletters}
\label{sphersymconstraints2}
\begin{eqnarray}
        \dr\kr - 2\gamr\kr - \mrr &=& 0
     \\
        \dr\kt - \mrt &=& 0
     \\
        \mtr - {\grt\over 2Br}(\krtil - \kt) &=& 0
     \\
        \dr(\ln N) - \ar &=& 0
     \\
        \at - {\grt\over 2A^2Br}\ar &=& 0       
     \\     
        \dr\ar - \arr - \gamr\ar &=& 0
     \\
        \dr A - A\gamr &=& 0
     \\
        \dr Br - {\grt\over 2} &=& 0.  
\end{eqnarray}
\end{mathletters}

\section{Causal Evolution Method} \label{sec:evolution}

Here we present the causal differencing method we use to evolve
Eqs.~(\ref{genevolution}) from one spacelike hypersurface to the next.
Straightforward differencing schemes typically become unstable for
large shifts, which are needed for the implementation of AHBC methods.
Our method is second-order accurate and has a stability criterion that
is independent of the shift vector. We emphasize that our method is
not specific to Eqs.~(\ref{genevolution}), but can be used to handle
advective terms in any system of first-order evolution equations.
Similar causal differencing methods have been used by
\cite{seidel_suen92,anninos_etal95} and \cite{alcubierre94} in order
to treat large shifts in the standard ADM formulation of Einstein's
equations.

Our causal differencing method is independent of the actual form of
the shift vector. We only place two restrictions on the shift: First,
it must be a smooth functional of the dynamical variables and of the
space and time coordinates. Second, it can be computed to second-order
accuracy, given second-order values for these variables and
coordinates.  The particular prescription that we use for computing
the shift will be discussed in section~\ref{sec:shift}.  In this
section we assume only that such a prescription exists.

Although the code described in this paper assumes spherical symmetry,
our causal evolution method is general. In this section we first
describe our method for the general case of three spatial dimensions
plus time, and then we specialize to spherical symmetry.

\subsection{Overview of the Method} \label{sec:evolution_basic}

\begin{figure}[t] 
\begin{center}
\begin{picture}(250,250)
\put(0,0){\epsfxsize=3.5in\epsffile{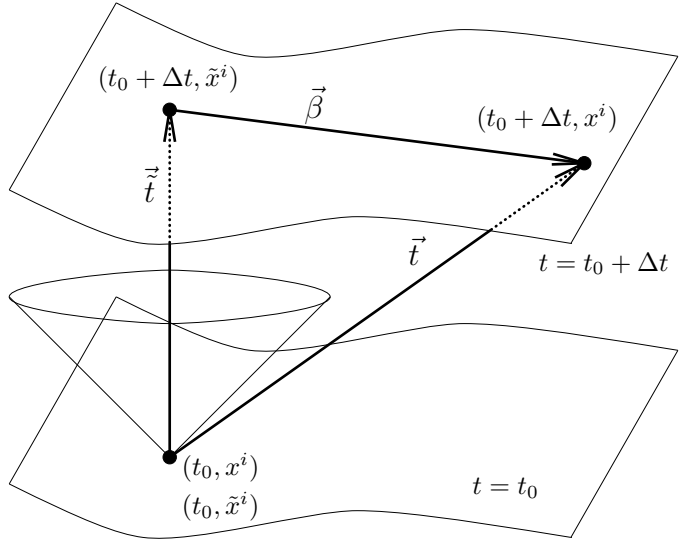}}
\put(65,28){\makebox(0,0)[l]{$(t_0,x^i)$}}
\put(65,13){\makebox(0,0)[l]{$(t_0,\tilde{x}^i)$}}
\put(60,175){\makebox(0,0)[c]{$(t_0+\Delta t,\tilde{x}^i)$}}
\put(230,160){\makebox(0,0)[r]{$(t_0+\Delta t,x^i)$}}
\put(115,165){\makebox(0,0)[c]{\large $\vec\beta$}}
\put(55,135){\makebox(0,0)[r]{\large $\vec{\tilde{t}}$}}
\put(155,110){\makebox(0,0)[r]{\large $\vec t$}}
\put(200,20){\makebox(0,0)[r]{$t=t_0$}}
\put(200,105){\makebox(0,0)[l]{$t=t_0+\Delta t$}}
\end{picture}
\end{center}
\caption{Spacetime diagram illustrating the relation
$\vec{t}=\vec{\tilde{t}}+\vec{\beta}$ and showing both the $(t,x^i)$
and the $(\tilde{t},\tilde{x}^i)$ coordinate systems.  The vector
$\vec{\tilde{t}}$ must always lie within the light cone. This is not
true for the vector $\vec{t}$ for a sufficiently large shift
$\vec{\beta}$.}
\label{fig:spacetime}
\end{figure}

Figure~\ref{fig:spacetime} shows an initial spatial hypersurface
labeled by $t=t_0$, and a subsequent spatial hypersurface labeled by
$t=t_0+\Delta t$. We wish to evolve quantities defined at the point
$(t_0,x^i)$ to the point $(t_0+\Delta t,x^i)$, that is, along the
vector $\vec{t}$ shown in Figure~\ref{fig:spacetime}.  Here $t$ and
$x^i$ are the coordinates defined in Eq.~(\ref{fourmetric}), and the
vector $\vec{t}$ is given by
\begin{equation}
\vec{t} = N\vec{n} + \vec{\beta},
\end{equation}
where $\vec{n}$ is the unit normal to the hypersurface $t=t_0$.

A large shift vector $\vec\beta$ tends to cause stability problems in
most numerical schemes. Some schemes, including many implicit ones,
are unconditionally unstable whenever $\vec{t}$ is non-timelike, as is
the case in Figure~\ref{fig:spacetime}. Other schemes can be made
stable for an arbitrary shift, but only at the expense of a very small
time step $\Delta t$.

In order to construct a differencing scheme that works for an
arbitrary shift, we introduce an auxiliary coordinate system.  First
define a new timelike vector
\begin{equation} \label{ttildedef}
\vec{\tilde{t}}\equiv N\!\vec{n} \equiv \vec{t} -\vec{\beta}.
\end{equation}
Then define new coordinates $(\tilde{t},\tilde{x}^i)$ such that
\begin{eqnarray}
\tilde{t} &=& t\\
\tilde{x}^i &=& \tilde{x}^i(x^i,t)\\
\pounds_{\tilde t} \tilde{x}^i &=& 0,
\end{eqnarray}
and such that the spatial coordinates ${\tilde{x}^i}$ coincide with
$x^i$ at $t=t_0$.  The new coordinates and their relationship to the
vectors $\vec{t}$ and $\vec{\tilde{t}}$ are shown in
Figure~\ref{fig:spacetime}.  Partial derivatives with respect to the
new coordinates $(\tilde{t},\tilde{x}^i)$ are given by
\begin{eqnarray}
{\partial\over\partial \tilde{t}} &=& 
{\partial\over\partial t}
-\beta^i{\partial \over \partial x^i}, \label{dbydttildef}\\
{\partial\over\partial \tilde{x}^i} &=&
{\partial x^j\over \partial \tilde{x}^i}
{\partial\over\partial x^j}. \label{dbydxtildef}
\end{eqnarray}

\begin{figure}[t] 
\begin{center}
\begin{picture}(250,250)(-25,0)
\put(0,0){\epsfxsize=3.0in\epsffile{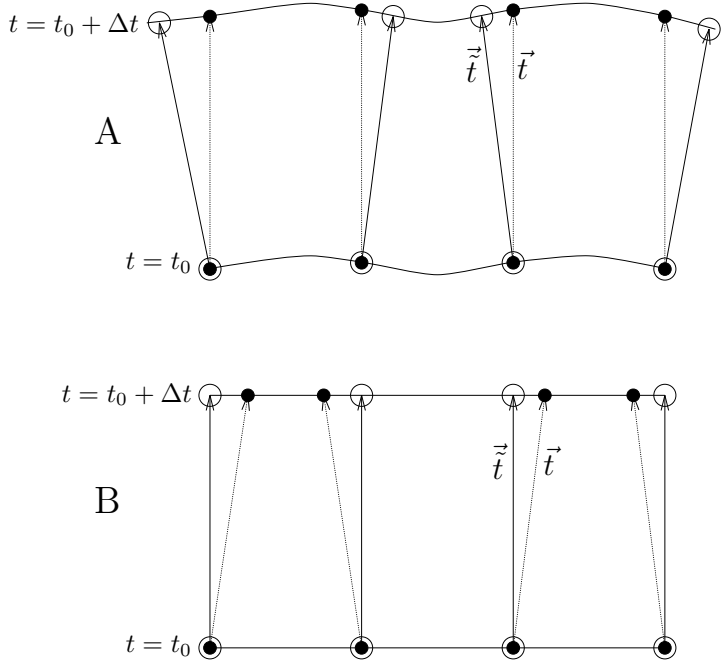}}
\put(17,5){\makebox(0,0)[r]{ $t=t_0$}}
\put(17,150){\makebox(0,0)[r]{$t=t_0$}}
\put(17,100){\makebox(0,0)[r]{$t=t_0+\Delta t$}}
\put(-3,240){\makebox(0,0)[r]{$t=t_0+\Delta t$}}
\put(135,75){\makebox(0,0)[r]{\large  $\vec{\tilde{t}}$}}
\put(150,75){\makebox(0,0)[l]{\large  $\vec{t}$}}
\put(125,225){\makebox(0,0)[r]{\large  $\vec{\tilde{t}}$}}
\put(140,225){\makebox(0,0)[l]{\large  $\vec{t}$}}
\put(-20,60){\makebox(0,0)[l]{\Large B}}
\put(-20,200){\makebox(0,0)[l]{\Large A}}
\end{picture}
\end{center}
\caption{Two-dimensional spacetime diagrams illustrating the two
coordinate systems used in our causal differencing scheme.  Solid and
dotted arrows denote the normal vector $\vec{\tilde{t}}$ and the time
vector $\vec{t}$ at each grid point on the initial time slice.  Solid
dots represent grid points with particular values of $x^i$, and
circles represent grid points with particular values of $\tilde{x}^i$.
The dots and circles coincide at $t=t_0$, but not at $t=t_0+\Delta t$.
This is because $x^i$ is constant along $\vec{t}$ while $\tilde{x}^i$
is constant along $\vec{\tilde{t}}$.  Diagrams A and B represent the
same spacetime. The only difference is that A is drawn in the
computational coordinate system $(t,x^i)$, where the time axis lies
along $\vec{t}$, and B is drawn in the normal coordinate system
$(\tilde{t},\tilde{x}^i)$, where the time axis lies along
$\vec{\tilde{t}}$ and is normal to the spatial slices.}
\label{fig:coordsystems}
\end{figure}

Our method works by breaking up each time step into two substeps, as
illustrated in Figure~\ref{fig:coordsystems}: First, we evolve
quantities along the vector $\vec{\tilde{t}}$, that is, we evolve
using the $(\tilde{t},\tilde{x}^i)$ coordinate system, from the points
on the slice $t=t_0$ in Figure~\ref{fig:coordsystems} to the points on
the slice $t=t_0+\Delta t$ that are labeled by circles.  We then
complete the timestep by interpolating from the points labeled by
circles to those labeled by dots.  These two substeps will be
considered separately in sections~\ref{sec:macormack}
and~\ref{sec:evolution2} below.

\subsection{Transforming into the $(\tilde{t},\tilde{x}^i)$ system}
\label{sec:lierewrite}		

Each evolution equation in the system~(\ref{genevolution}) can be
written in the form
\begin{equation}\label{prototype}
\dt T - Q {\partial\over\partial x^i}S^i = R,
\end{equation}
where $T$, the quantity being evolved, is not necessarily a scalar,
but may be a coordinate-dependent object.  We wish to rewrite
Eq.~(\ref{prototype}), which is defined in the computational
coordinate system $(t,x^i)$, in terms of the normal coordinate system
$(\tilde{t},\tilde{x}^i)$.

If we consider all quantities to be defined {\it in the $(t,x^i)$
basis}, we can rewrite the $\dt$ operator in the
$(\tilde{t},\tilde{x}^i)$ coordinate system as follows:
\begin{eqnarray}\label{lierewrite}
\dt
        &=& \pounds_{\tilde{t}} \nonumber \\
        &=& \pounds_{t}-\pounds_{\beta} \nonumber \\
        &=& {\partial\over \partial t}
                 - \pounds_{\beta} \nonumber \\
        &=& {\partial\over \partial t}
                 - \beta^i{\partial \over \partial x^i}
                        - (\pounds_{\beta} 
	                  - \beta^i{\partial\over \partial x^i}) \nonumber \\
        &=& {\partial\over\partial\tilde{t}}
	                - (\pounds_{\beta} 
	                  - \beta^i{\partial\over \partial x^i}),
\end{eqnarray}
so that 
Eq.~(\ref{prototype}) becomes
\begin{equation}\label{rewriteprototype}
{\partial\over\partial\tilde{t}}  T 
- Q {\partial \tilde{x}^j\over\partial x^i}
   {\partial\over\partial\tilde{x}^j}S^i 
= {\hat{\mathcal L}}_{\beta}T + R,
\end{equation}
where
\begin{eqnarray}
{\hat{\mathcal L}}_{\beta}&\equiv& \pounds_{\beta}
- \beta^i{\partial\over\partial x^i}\\
&=&
\pounds_{\beta}
- \beta^i
{\partial \tilde{x}^j\over\partial x^i}
{\partial\over\partial\tilde{x}^j}.
\label{definelbeta}
\end{eqnarray}

Here we have used Eqs.~(\ref{dt}), (\ref{ttildedef}), and
(\ref{dbydttildef}). The first two lines of Eq.~(\ref{lierewrite}) are
coordinate independent, but in the third line we have assumed the
$(t,x^i)$ basis in order to write $\pounds_t =\partial/\partial t$.
In the fourth line we have separated the Lie derivative along
$\vec\beta$ into two pieces: The advective piece,
$\beta^i\partial/\partial x^i$, is the one responsible for the
instability that often arises when one tries to evolve along $\vec{t}$
with a large shift vector.  This piece is eventually absorbed into the
time derivative $\partial/\partial\tilde{t}$.  The remaining piece,
${\hat{\mathcal L}}_{\beta}$, when operating on some quantity $T$,
describes the change in $T$ induced by the change in basis vectors
along $\vec\beta$.  The operator ${\hat{\mathcal L}}_{\beta}$ vanishes
when operating on a scalar.  Furthermore, ${\hat{\mathcal L}}_{\beta}
T$ does not actually contain any derivatives of $T$, but only contains
derivatives of $\vec{\beta}$.  Therefore, no spatial derivatives of
$T$ appear in Eq.~(\ref{rewriteprototype}).

Note that Eq.~(\ref{rewriteprototype}) is not the same as
Eq.~(\ref{prototype}) transformed into the $(\tilde{t},\tilde{x}^i)$
basis.  By splitting the Lie derivative along $\vec\beta$ into two
pieces, we have derived an equation that describes the evolution of
$T$ {\it defined with respect to the $(t,x^i)$ basis\/} along a path
of constant $\tilde{x}^i$.  The coordinate system used to define
tensor components, $(t,x^i)$, is different from the coordinate system
used to label spacetime points during the evolution,
$(\tilde{t},\tilde{x}^i)$.

Note also that we have introduced an additional auxiliary variable:
the Jacobian $\partial \tilde{x}^j/\partial x^i$ that appears in
Equation~(\ref{rewriteprototype}).  From Eqs.~(\ref{dbydttildef})
and~(\ref{dbydxtildef}), we can find the rate of change of the
Jacobian along the vector $\vec{\tilde{t}}$:
\begin{equation} \label{evolvejacoby}
{\partial\over\partial\tilde{t}}
\left[\partial \tilde{x}^i\over\partial x^j\right] =
{\partial \tilde{x}^i\over\partial x^k}
{\partial \tilde{x}^\ell\over\partial x^j}
{\partial \beta^k\over\partial \tilde{x}^\ell}.
\end{equation}
To compute the Jacobian, we evolve this equation along with
Eq.~(\ref{rewriteprototype}). Because we set $\tilde{x}^i=x^i$ at the
beginning of each time step $t=t_0$, the initial value of $\partial
\tilde{x}^j/\partial x^i$ on each time step is the Kroeneker delta
$\delta^j_i$.

Finally, we require derivatives of $\vec{\beta}$, which appear in the
operator ${\hat{\mathcal L}}_{\beta}$ and also on the right hand side
of Eq.~(\ref{evolvejacoby}).  Assuming that we have some prescription
for choosing the shift given the values of $x^i$, $t$, and the
dynamical variables at each grid point, we simply use this
prescription to compute $\vec\beta$, and then we obtain its
derivatives either analytically (if the shift is analytic) or by
finite differencing.

\subsection{Step 1: Evolve along $\vec{\tilde{t}}$} \label{sec:macormack}

The first step of our causal differencing method is to evolve
Eq.~(\ref{rewriteprototype}) and the auxiliary
equation~(\ref{evolvejacoby}) along the vector $\vec{\tilde{t}}$, from
$\tilde{t}=t_0$ to $\tilde{t}=t_0+\Delta t$. Although this can be done
using any standard differencing scheme, the algorithm described in
this section assumes a scheme with two time levels; a three-level
scheme (such as leapfrog) requires a slight modification of the
algorithm. We use a Macormack predictor-corrector method, which we
illustrate with a simple wave equation in spherical symmetry, written
in first-order form in the $(\tilde{t},\tilde{r})$ coordinate system:
\begin{eqnarray} \label{simplewaveequation}
{\partial\over\partial \tilde{t}}P - 
{\partial\over\partial \tilde{r}}Q = R \\
{\partial\over\partial \tilde{t}}Q -
{\partial\over\partial \tilde{r}}P = S.
\end{eqnarray}
Here $R$ and $S$ are arbitrary functions of $P$, $Q$, $\tilde{r}$, and
$\tilde{t}$.  Given a discrete set of uniformly spaced grid points
$\tilde{r}_i$, we denote $P$ and $Q$ at grid point $i$ and time
$\tilde{t}=\tilde{t}_n$ by $P^n_i$ and $Q^n_i$ respectively.  To
compute $P$ and $Q$ at time
$\tilde{t}=\tilde{t}_{n+1}=\tilde{t}_n+\Delta \tilde{t}$, we first
compute initial guesses $\bar P$ and $\bar Q$ using the ``predictor''
step
\begin{eqnarray}
\bar{P}^{n+1}_i &=& P^n_i 
	+ {\Delta \tilde{t}\over \Delta \tilde{r}}(Q^n_{i+1} - Q^n_i)
	+ R(P^n_i,Q^n_i) \label{macormackwavep1}\\
\bar{Q}^{n+1}_i &=& Q^n_i 
	+ {\Delta \tilde{t}\over \Delta \tilde{r}}(P^n_{i+1} - P^n_i)
	+ S(P^n_i,Q^n_i), \label{macormackwavep2}
\end{eqnarray}
where $\Delta \tilde{r}=\tilde{r}_{i+1}-\tilde{r}_{i}$.
The quantities $\bar{P}^{n+1}_i$ and $\bar{Q}^{n+1}_i$ are first
order accurate in both space and time. Notice that the finite
difference approximation to the spatial derivative is one-sided.

Once we have the predicted quantities, we then compute
$P$ and $Q$ at time
$\tilde{t}=\tilde{t}_{n+1}=\tilde{t}_n+\Delta \tilde{t}$
to second order in both
space and time using the ``corrector'' step
\begin{eqnarray}
P^{n+1}_i &=& {1\over 2}\left(
         \bar{P}^{n+1}_i + P^n_i + {\Delta t\over \Delta \tilde{r}}
	 (\bar{Q}^{n+1}_i - \bar{Q}^{n+1}_{i-1})
	+ R(\bar{P}^{n+1}_i,\bar{Q}^{n+1}_i)
	\right) \label{macormackwavec1}\\
Q^{n+1}_i &=& {1\over 2}\left(
         \bar{Q}^{n+1}_i + Q^n_i + {\Delta t\over \Delta \tilde{r}}
	 (\bar{P}^{n+1}_i - \bar{P}^{n+1}_{i-1})
	+ S(P^{n+1}_i,\bar{Q}^{n+1}_i)
	\right). \label{macormackwavec2}
\end{eqnarray}
The spatial derivatives are taken in the opposite direction to the
predictor step. This ensures that the first order error terms
introduced by the one-sided derivatives in the corrector step cancel
those produced by the one-sided derivatives in the predictor step.
This cancellation would be spoiled by substituting $Q^{n+1}$ for
$\bar{Q}^{n+1}$ in the spatial derivative term of
Eq.~(\ref{macormackwavec1}) or by substituting $P^{n+1}$ for
$\bar{P}^{n+1}$ in the spatial derivative term of
Eq.~(\ref{macormackwavec2}). However, it is irrelevant for accuracy or
stability whether the right hand sides $R$ and $S$ in the corrector
step are computed using predicted values of $P$ and $Q$ as in
Eq.~(\ref{macormackwavec1}) or using corrected values of $P$ and
predicted values of $Q$ as in Eq.~(\ref{macormackwavec2}).  We use
corrected values in the right-hand sides whenever possible because it
minimizes the memory required for storing temporary variables on the
computer.

The above Macormack scheme is stable (disregarding the boundaries,
which will be discussed in section~\ref{sec:outerbdry}) whenever
$\Delta \tilde{t} < \Delta \tilde{r}$. For a wave equation with
characteristic speed $v$, the stability condition for the above scheme
is the familiar Courant condition $v\Delta \tilde{t} < \Delta
\tilde{r}$.

To obtain a corrected value for a particular variable, we require
predicted values for all quantities appearing in the equation for that
variable.  For our prototypical equation~(\ref{rewriteprototype}), in
order to compute a corrected value for $T$, we must have predicted
values not only for $T$, $Q$, $S^i$, and $R$, but also for the
Jacobian $\partial \tilde{x}^j/\partial x^i$ and for $\vec\beta$ and
its derivatives (which appear in the operator ${\hat{\mathcal L}}_{
\beta}$).  The predicted value of the Jacobian is obtained by evolving
Eq.~(\ref{evolvejacoby}) to first order using the Macormack predictor
step.  The predicted value of the shift is obtained by using our shift
prescription to compute $\vec{\beta}$ from the predicted values of the
dynamical variables.

\subsection{Step 2: Interpolation} \label{sec:evolution2}

One entire numerical time step should take variables defined on
discrete points with particular values of $x^i$, and compute
quantities at the {\it same\/} values of $x^i$ but at a later time.
In Figure~\ref{fig:coordsystems}, this corresponds to taking values
defined at the discrete points on the slice $t=t_n$ and computing
quantities at the solid dots on the slice $t=t_n+\Delta t$.  However,
when we evolve along $\vec{\tilde{t}}$ as described in
section~\ref{sec:macormack}, we compute quantities at the points that
correspond to the circles on the slice $t=t_n+\Delta t$ in
Figure~\ref{fig:coordsystems}.

Our causal differencing method therefore requires a second step,
namely, interpolation of our dynamical variables from the
$(\tilde{t},\tilde{x}^i)$ coordinate system back into the $(t,x^i)$
coordinate system, that is, from the circled points to the dotted
points on the slice $t=t_n+\Delta t$ in Figure~\ref{fig:coordsystems}.
The values of $x^i$ at the dotted points and the values of
$\tilde{x}^i$ at the circled points are known; both are the same as
the values of $x^i$ at the appropriate grid points on the initial
slice.  To perform the interpolation, we must also know either the
values of $x^i$ at the circled points or the values of $\tilde{x}^i$
at the dotted ones.

From Eq.~(\ref{dbydttildef}), the change in the coordinate $x^i$ along
the vector $\vec{\tilde{t}}$ is given by
\begin{equation}\label{evolvex}
{\partial x^i\over\partial\tilde{t}} = -\beta^i.
\end{equation}
Therefore, if we evolve Eq.~(\ref{evolvex}) along with
Eq.~(\ref{rewriteprototype}) using the Macormack scheme, we obtain the
value of $x^i$ at each of the circled grid points at $t=t_n+\Delta t$
in Figure~\ref{fig:coordsystems}.  This allows us to interpolate
quantities from the circles to the dots, working in the $x^i$
coordinate system.

Note, however, that the circled grid points are, in general, not
uniformly spaced in $x^i$, as shown in
Figure~\ref{fig:coordsystems}A. Instead, the dotted grid points are
uniformly spaced in $x^i$.  While this poses no difficulty for the
spherically symmetric case discussed in this paper, in the general
three-dimensional case interpolating from an arbitrary set of points
onto a uniform grid is a nontrivial numerical problem that cannot be
treated very efficiently.  Much easier and much less costly in terms
of computer time is to interpolate from a uniform grid to an arbitrary
set of points.

To handle this difficulty, notice that if we evolve along the vector
$\vec t$ instead of along the vector $\vec{\tilde{t}}$, the
coordinates $x^i$ remain constant and the coordinates $\tilde{x}^i$
vary.  The change in $\tilde{x}^i$ along the vector $\vec t$ is given
by
\begin{equation}\label{evolvetildex}
{\partial\tilde{x}^i\over\partial t} =
{\partial\tilde{x}^i\over\partial x^j}
\beta^j,
\end{equation}
where we have used Eq.~(\ref{dbydttildef}).  Therefore, if we evolve
Eq.~(\ref{evolvetildex}) along the vector $\vec t$ using the Macormack
scheme, we obtain the value of $\tilde{x}^i$ at each of the dotted
grid points at $t=t_n+\Delta t$ in Figure~\ref{fig:coordsystems}.
This allows us to interpolate quantities from the circled points to
the dotted ones, working in the $\tilde{x}^i$ coordinate system.  The
circled points are uniformly spaced in $\tilde{x}^i$, as shown in
Figure~\ref{fig:coordsystems}B.  We thus interpolate from a uniform
grid (in $\tilde{x}^i$) to an arbitrary set of points. This can be
done relatively easily in three spatial dimensions.

There is a subtlety in evolving Eq.~(\ref{evolvetildex}) along the
vector $\vec t$ using the Macormack scheme: In order to implement the
corrector step of Eq.~(\ref{evolvetildex}), we require predicted
values of the quantities $\beta^j{\partial\tilde{x}^i/\partial x^j}$
{\it at the dotted points\/}, but these quantities are only known at
the circled points.  However, from the predictor step of
Eq.~(\ref{evolvetildex}), we already have {\it predicted\/} values of
$\tilde x^i$ at the dotted points.  Therefore, we use these values to
interpolate the predicted values of
$\beta^j{\partial\tilde{x}^i/\partial x^j}$ from the circled points to
the dotted ones.  In this case, we are again interpolating from a
uniform grid to an arbitrary set of points.

\subsection{Implementation in Spherical Symmetry} \label{sec:evolution3}

In spherical symmetry, we solve equations~(\ref{CBYespher}) on a
numerical grid that extends from some value $r=r_{min}$ just inside
the apparent horizon to a large radius $r=r_{max}$.  We denote our
numerical grid points by $r_i$, where $i$ runs from $i_{min}$ to
$i_{max}$, corresponding to the innermost and outermost grid points.
We use a zone-centered grid, so that the innermost and outermost grid
points do not correspond to $r_{min}$ and $r_{max}$. Instead, we set
\begin{equation}
r_i = r_{min} + \left(i-i_{min}+{1\over 2}\right)\Delta r,
\end{equation}
where 
\begin{equation}
\Delta r = {r_{max}-r_{min}\over i_{max}-i_{min}+1}.
\end{equation}

When evolving along $\vec{\tilde{t}}$, the spherically symmetric
vacuum evolution equations~(\ref{CBYespher}) take the form
\begin{mathletters} \label{rewriteevolution}
\begin{eqnarray}
\dtt A    &=& -NA\krtil 
              + A\drr\beta^r \\
\dtt Br   &=& -NBr\kt\\
\dtt \kr  &=& \phantom{-}N\lr 
              +2\kr\drr\beta^r \\
\dtt \kt  &=& \phantom{-}N(\lt+2\kt^2)\\
\dtt N    &=& -N^2(\krtil+2\kt)\\
\dtt a_r  &=& \phantom{-}N\aor 
              +\ar\drr\beta^r \\
\dtt \at  &=& \phantom{-}N\left[(2\mtr-\mrt-\ar\kt){\ar\over A^2}
                     \nonumber \right. \\ &&\qquad\left.
              +{\grt\aor\over 2A^2Br}
              +2\kt\at\right]\\
\dtt \gamr &=& -{N\over A^2}\left[\kr\ar+\mrr\right]
                    \nonumber \\ &&\qquad
	      {}+\gamr\drr\beta^r
                +\drr\left(\drr\beta^r\right)
              \\
\dtt \grt &=& -N\left[\kt\grt+2Br(\ar\kt+\mrt)\right]
                    \nonumber \\ &&\qquad
              {}+\grt\drr\beta^r \\
\dtt \mtr &=& \phantom{-}N\left[
              2\mtr(\kt+\krtil)
              +(\mrt+\ar\kt)(\kt-\krtil) 
      \vphantom{\grt\over 2Br} \nonumber \right. \\ &&\qquad\left.
              +{\grt\over 2Br}\left({\lr\over A^2}-\lt\right)
              \right]
              {}+\mtr\drr\beta^r
\\
\dtt \arr   -   N\drr\aor
                     &=&\phantom{-}N\left[-\gamr\aor
	                   +\ar(\ar\krtil+\mrr/A^2+\aor)\right]
                    \nonumber \\ && \qquad
                          {}+2\arr\drr\beta^r
                           \\
\dtt \aor   -   {N \over A^2}\drr\arr
                     &=&\phantom{-}N\left[Q_r-{2\gamr\arr\over A^2}
                           +{\grt\over Br}({\arr/A^2}-\at)\right]
                    \nonumber \\ &&\qquad
 			   {}+\aor\drr\beta^r
                           \\
\dtt \mrr   -   N\drr\lr 
                     &=&\phantom{-}N\left[(\ar-2\gamr)\lr
                           +2\krtil(\kr\ar+\mrr)\right]
                    \nonumber \\ &&\qquad
                          {}+3\mrr\drr\beta^r \\
\dtt \mrt   -   N\drr\lt
                     &=&\phantom{-}N\left[2\kt(\ar\kt+2\mrt)+\ar\lt\right]
                    \nonumber \\ &&\qquad
                           {}+\mrt\drr\beta^r \\
\dtt \lr    -   {N \over A^2}\drr\mrr
                     &=&\phantom{-}N\left[-J_{rr}-{3\gamr\mrr\over A^2}
                           +{\grt\over Br}
                             \left({\mrr\over A^2}-2\mtr\right)\right]
                    \nonumber \\ &&\qquad
			   {}+2\lr\drr\beta^r
                            \\
\dtt \lt    -   {N \over A^2}\drr\mrt
		     &=&\phantom{-}N\left[-J_{T}
			   -{\gamr\mrt\over A^2}
                           +{\grt\over A^2 Br}(\mrt+\mtr)\right],
\end{eqnarray}
\end{mathletters}
where the right hand
sides~(\ref{right_hand_sides_begin}--\ref{right_hand_sides_end}) are
unchanged, and we have included the ${\hat{\mathcal L}}_{\beta}$ terms
explicitly.  Note the second derivative of $\beta^r$ in the equation
for $\gamr$. This term results from applying $\pounds_{\beta}$ to a
non-tensorial quantity. Because $\beta^r$ is not an unknown in the
system of equations~(\ref{rewriteevolution}), but is instead an
auxiliary variable that may be, for example, given analytically as a
function of the coordinates, this second derivative should not spoil
the hyperbolicity of the system.

In spherical symmetry, 
Equations~(\ref{evolvejacoby}), (\ref{evolvex}), and~(\ref{evolvetildex})
become
\begin{mathletters}\label{spherauxeqns}
\begin{eqnarray}
{\partial\tilde{r}\over\partial t} &=&
{\partial\tilde{r}\over\partial r}
\beta^r \label{spherauxeqn1}\\
{\partial r\over\partial\tilde{t}} &=& -\beta^r
\label{spherauxeqn2}\\
{\partial\over\partial\tilde{t}} 
\left[\partial \tilde{r}\over\partial r\right] &=&
\left(\partial \tilde{r}\over\partial r\right)^2
{\partial \beta^r\over\partial \tilde{r}}.
\label{spherauxeqn3}
\end{eqnarray}
\end{mathletters}
Eqs.~(\ref{spherauxeqns})
provide values for ${\partial \tilde{r}/\partial r}$ to be used in the
corrector step, as well as coordinate information for the
interpolation step.

We use cubic interpolation for causal differencing. This is accurate
to fourth order in $\Delta \tilde{r}$.  Quadratic interpolation would
also suffice, but linear interpolation would not yield a scheme that
was second order convergent in time.  The reason is that linear
interpolation makes an ${\mathcal O}(\Delta \tilde{r})^2$ error {\it on
each time step\/}, so that after $N$ time steps the error is of order
$\Delta\tilde{r}$. This is because for a fixed total time $t_{\rm
total}$ that we wish to evolve, the Courant condition requires $N\sim
t_{\rm total}(\Delta\tilde{r})^{-1}$.

\section{Shift Vector and Inner Boundary Condition} \label{sec:shift}

In this section we describe how we choose a shift vector
$\vec{\beta}$, and how this choice affects how we handle the inner
boundary of our computational domain.

Because both the CBY formalism and the causal differencing scheme
discussed in section~\ref{sec:evolution} place no restrictions on
$\vec{\beta}$, we are free to choose any shift we wish.  Although
setting $\vec{\beta}=0$ is the simplest choice, it is often useful to
employ a nonzero shift vector.  One technique is to use the shift to
simplify the form of the Einstein equations, for example, to eliminate
particular components of the spatial metric \cite{st85}.  The
disadvantage of this approach is that it involves an actual change in
the equations being solved. Instead of Eqs.~(\ref{genevolution}), one
would be solving a different (but physically equivalent) system of
equations that would include the shift as a dynamical variable, and
would no longer be hyperbolic.

We instead use the shift for a different purpose: to allow us to
truncate our computational domain just inside the apparent horizon, so
that we evolve only the exterior region.  We thus avoid the spacetime
singularity inside the black hole.  If we were to attempt such a
truncation with $\vec{\beta}=0$, numerical grid points originally
located just outside the apparent horizon would soon fall in, and any
grid points located inside the apparent horizon would eventually
encounter the singularity. This is because for zero shift, numerical
grid points follow the world lines of normal observers, and these
world lines are necessarily timelike.

\subsection{Shift at the Inner Boundary} \label{sec:shift_AH}

We force the inner boundary of our grid, $r=r_{min}$, to hover within
a grid spacing of the apparent horizon.  To accomplish this for a
static black hole, we choose $\vec\beta$ near the horizon to point
along the outward normal to the horizon, and we choose its magnitude
so that the local coordinate speed of light in that direction is zero.
The horizon is then approximately stationary with respect to the
spatial coordinates.  For the spherically symmetric case, the local
coordinate speed of light in the outgoing radial direction is given by
\begin{equation}
\label{speedoflight}
c={dr\over dt} = {N\over A} -\beta^r,
\end{equation}
so we set $\beta^r$ at the horizon equal to $N/A$.

We track the apparent horizon at each time step, and retain only the
grid points that lie on the outside. This is done via a masking
algorithm that labels grid points outside the apparent horizon as
valid, and those inside as invalid.  Invalid points are never used in
the computation, just as if those points did not exist. Because we use
a cell-centered grid, the inner boundary $r=r_{min}$ is located half a
grid spacing inside the innermost valid grid point.

Numerical errors typically cause the horizon to drift through the
grid, even if one tries to lock it in place by forcing $\beta^r =N/A$.
While this drift seems to cause minimal difficulty with either
the stability or accuracy of the code, for coarse-resolution runs it
produces a small but distracting gauge pulse each time the horizon
crosses a grid zone.  Horizon drift can be eliminated by introducing a
feedback mechanism\cite{seidel_suen_private} that adjusts the
magnitude of the shift to compensate.  We do this as follows: If we
wish to force the horizon to remain at some radius $r_0$, but we find
the horizon is actually at radius $r_{AH}$, then we set
\begin{equation}
\beta^r = {N\over A} + {(r_{AH}-r_0)\over\Delta t},
\end{equation}
where $\beta^r$, $N$, and $A$ refer to values at the horizon location
$r_{AH}$. This feedback mechanism is not necessary for sufficiently
fine grid spacing.

While locating an apparent horizon on a numerically generated time
slice is a difficult problem in
multidimensions\cite{nakamura84,libson95,huq96,baumgarte_etal96}, it
is trivial in spherical symmetry.  The marginal outer trapped surface
equation
\begin{equation} \label{AHeqn}
D_is^i+s^is^jK_{ij}-K^i_i=0,
\end{equation}
where $s^i$ is the spatial unit normal to the surface,
reduces to
\begin{equation} \label{AHeqnspher}
\vartheta(r) = \grt/A - 2Br\kt = 0
\end{equation}
for a spherically symmetric system. Here we have used our variables
defined in Eqs.~(\ref{3metric}) and~(\ref{sphericalvariables}).  We
find the apparent horizon by first evaluating $\vartheta(r)$ at each
grid point, and then by using 3-point interpolation to locate the
outermost root that satisfies $\vartheta'(r)>0$.  When locating the
apparent horizon after the Macormack predictor step, we must be sure
to use the value of $r$ and not $\tilde{r}$ in evaluating the function
$\vartheta(r)$ at each grid point.

In principle, one can also use a shift condition at the apparent
horizon to move a black hole through the numerical grid.  This could
be accomplished by adjusting the shift so that grid points on one side
of the hole fall into the horizon, and grid points on the other side
emerge from it.  A similar idea could in principle be applied to
rotating black holes or systems with more than one black hole.

\subsection{Inner Boundary Condition on Eqs.~(\ref{CBYespher})}
\label{sec:shift_IBC}

Treating the inner boundary correctly is a primary motivation for
using a hyperbolic formulation of the Einstein equations.  The
spherically symmetric CBY equations have only three characteristics at
each spacetime point. These lie along the ingoing and outgoing null
rays, and along the normal vector $\vec{\tilde{t}}$.  A boundary
condition is required only at a point that cannot obtain information
from one or more of the characteristics passing through it. For
example, the outer boundary cannot obtain information from the ingoing
null characteristics that it intersects, because these characteristics
originate from outside the computational domain, where we have no
data. To update quantities at the outer boundary, this information
must be provided by a boundary condition.  Similarly, the inner
boundary ordinarily requires a boundary condition because it cannot
obtain information from the outgoing null characteristics that it
intersects.  However, if the inner boundary follows an outgoing
spacelike or null trajectory, then each point on the inner boundary
can obtain information from all three characteristics passing through
it, so a boundary condition is not required.

This is the essence of an AHBC scheme for treating the inner boundary:
by forcing the inner boundary to move along with the apparent horizon,
we force it along an outgoing non-timelike path. Therefore, there is
no need to impose an explicit inner boundary condition.  Regardless of
the mathematical formulation of Einstein's equations being used,
general relativity tells us that when the inner boundary follows an
outgoing spacelike or null path, information with physical content
cannot penetrate this boundary from the inside, since this information
cannot propagate outside the light cone.  A key advantage of a
hyperbolic formulation with only simple (non-spacelike)
characteristics is that in such a formulation, this statement applies
to gauge information as well.

The way we solve Eqs.~(\ref{CBYespher}) without imposing an explicit
condition at the inner boundary is by simply ignoring the innermost
grid point during the interpolation step of our causal differencing
scheme.  In the case where the inner boundary $r=r_{min}$ moves with
respect to the $(\tilde{t},\tilde{r})$ coordinate a distance less than
$\Delta \tilde{r}$ during each time step, the interpolation becomes an
extrapolation at the innermost point.  This is always the case in our
simulation because of the Courant limit: Because the inner boundary,
which moves along with the apparent horizon, has a velocity with
respect to the $(\tilde{t},\tilde{r})$ coordinate system of
approximately $N/A$, the inner boundary can never move farther than
$\Delta \tilde{r}$ during a time step without violating the Courant
condition $\Delta t < (A/N) \Delta \tilde{r}$.  One could avoid this
extrapolation by using an implicit differencing
scheme\cite{alcubierre94,scheel95a} to get around the Courant
condition, but such a scheme requires much more computer time than
explicit schemes, especially in the multidimensional case.

\subsection{Shift in the Remainder of the Spacetime}
\label{sec:shift_elsewhere}

Once a shift criterion at the apparent horizon has been chosen, one
must then determine the shift in the remainder of the spacetime.  At
spatial infinity, one presumably would like the shift to approach
zero, so that the spacetime metric components approach Minkowski
values.  Or perhaps, in the case of spacetimes with nonzero angular
momentum, one would like the asymptotic shift to describe a
co-rotating frame.  However, given the shift at infinity and at the
apparent horizon, it is not clear how to choose the shift elsewhere.

One possibility is to use a parameterized analytic function whose
parameters are set so that the shift behaves appropriately near the
apparent horizon and far from the black hole.  For example, when
evolving a single black hole in spherical symmetry we have tried the
Gaussian form
\begin{equation} \label{gaussianshift}
\beta^r = C e^{-(r-r_c)^2/w^2},
\end{equation}
where $C$, $r_c$, and $w$ are chosen so that $\beta^r$ is equal to
$N/A$ at the apparent horizon, $\partial\beta^r/\partial r$ is equal
to $\partial(N/A)/\partial r$ at the apparent horizon, and $\beta^r$
is smaller than some threshold at the outer boundary of the grid.
Although this choice results in a second order convergent evolution,
we find that the grid points tend to compress or stretch where the
shift gradients are large, and eventually coordinate singularities
develop that cause the simulation to terminate.  Similarly, one can
choose
\begin{equation} \label{tanhshift}
\beta^r = {N\over 2A} (1-\tanh((r-r_c)/w)),
\end{equation}
so that the shift is equal to $N/A$ far inside some arbitrary radius
$r=r_c$, and zero far outside $r=r_c$.  In this case, the grid points
become compressed near $r=r_c$ as one evolves in time, and again the
simulation terminates.

It therefore appears that in addition to a prescription for specifying
the shift at the apparent horizon and at infinity, one must impose
some additional restriction on the shift that ensures that it will not
induce any coordinate pathologies elsewhere.  Such a restriction can
be provided by two different elliptic shift conditions that were
introduced for the very purpose of minimizing coordinate strain caused
by a shift vector.  The first is the minimal distortion condition
\cite{smarryork78}, which can be written in the form
\begin{equation} \label{MinDist}
D^jD_j\beta^i + {1\over 3} D^iD_j\beta^j + \bar{R}^i_j\beta^j
= 2D^j\left[ N \left(
K^i_j - {1\over 3} g^i_jK^k_k\right)\right].
\end{equation}
The minimal distortion shift minimizes the average change of shape of
a spatial volume element as it is dragged from one time slice to the
next.  A related choice is the minimal strain condition
\cite{smarryork78}, which can be written in the form
\begin{equation} \label{MinStrain}
D^jD_j\beta^i + D^iD_j\beta^j + \bar{R}^i_j\beta^j
= 2D^j\left(NK^i_j\right).
\end{equation}
The minimal strain shift minimizes the average change in the three
metric $g_{ij}$ as one evolves from one time slice to the next. It
differs from the minimal distortion shift in that it takes into
account the change in size of spatial volume elements as well as their
change in shape.

The downside of these shift conditions is that they require one to
solve elliptic equations. This can be costly in terms of computer
time, especially in three dimensions.  It may be possible to use a
parameterized analytic function to mimic one of these conditions, or
it may suffice to use an approximate solution. However, since it
appears that these conditions give us a useful shift vector, we will
adopt them in the spherically symmetric case, where the computational
burden is not so severe.

Both the minimal distortion and minimal strain conditions work well in
spherical symmetry, as shown in section~\ref{sec:tests}. They prevent
grid points from becoming locally compressed or stretched to the point
where coordinate singularities form.  Using the variables defined
in~(\ref{yorkvariables}) and (\ref{sphericalvariables}), we find that
the minimal distortion equation~(\ref{MinDist}) takes the form
\begin{eqnarray}
\left(\jacoby\right)^2{\partial^2\over\partial \tilde{r}^2}\beta^r
&+&\left(\gamr + {\grt\over Br}-\left(\jacoby\right)^2
{\partial^2 r \over\partial\tilde{r}^2}\right)\drr\beta^r \nonumber\\
&+&\beta^r\left(\drr + {3\over 2}{\grt\over Br}\right)
\left(\gamr - {\grt\over 2Br}\right)\nonumber\\
&=&N\left[\ar\left(\krtil-\kt\right) +{\mrr\over A^2}- \mrt +3\mtr\right],
 \label{MinDistSpher2}
\end{eqnarray}
and the minimal strain equation~(\ref{MinStrain}) becomes
\begin{eqnarray}
\left(\jacoby\right)^2{\partial^2\over\partial \tilde{r}^2}\beta^r
&+&\left(\gamr + {\grt\over Br}-\left(\jacoby\right)^2
{\partial^2 r \over\partial\tilde{r}^2}\right)\drr\beta^r \nonumber\\
&+&\beta^r\left(\drr\gamr
+ {\grt\over Br}
\left(\gamr - {\grt\over 2Br}\right)\right)\nonumber\\
&=&N\left[\ar\krtil +{\mrr\over A^2}+2\mtr\right].
\label{MinStrainSpher2}
\end{eqnarray}
We have written both equations in the $\tilde{r}$ coordinate system
and included the factors $\partial\tilde{r}/\partial r$ and
$\partial^2\tilde{r}/\partial r^2$ because the shift must be computed
not only after the corrector step of the Macormack scheme when
$\tilde{r}=r$, but also after the predictor step, when $\tilde{r}\neq r$.

Eqs.~(\ref{MinDistSpher2}) and~(\ref{MinStrainSpher2}) require boundary
conditions at both ends of the numerical grid.  We impose the
condition
\begin{equation} \label{shiftbcouter2}
\drr\beta^r+{n\beta^r\over r}= 0\qquad (n\ge 1)
\end{equation}
at the outer boundary $r=r_{max}$ so that the
spacetime is asymptotically Minkowski, and we impose
\begin{equation} \label{shiftbcinner}
\beta^r={N\over A}
\end{equation}
at the apparent horizon, so that the apparent horizon is stationary
with respect to the coordinates.

Both the minimal distortion and minimal strain equations
can be written in the general form
\begin{equation}\label{MinDistGeneral}
{\partial^2\over \partial\tilde{r}^2}\beta^r
+2Q{\partial\over \partial\tilde{r}}\beta^r+P\beta^r = R.
\end{equation}
We solve this equation using the usual
three-point finite difference approximation
\begin{equation} \label{MinDistFD}
{1\over (\Delta \tilde{r})^2}
\left(\beta^r_{i+1}-2\beta^r_{i}+\beta^r_{i-1}\right)
+{1\over \Delta \tilde{r}}
Q_i\left(\beta^r_{i+1}-\beta^r_{i-1}\right)
+P_i\beta^r_i = R_i.
\end{equation}

In order to retain second-order accuracy, we must be careful always to
impose the boundary condition~(\ref{shiftbcouter2}) at the point
$r=r_{max}$, which may be different from the outer boundary of the
grid, $\tilde{r}=r_{max}$.  We write Eq.~(\ref{shiftbcouter2}) in the
second-order accurate finite-difference form
\begin{eqnarray} \label{shiftbcouter2FD}
&&\left(\partial\tilde r\over \partial r\right)_{i}
{1\over \Delta \tilde{r}}\left(
(2\lambda+1)\beta^r_{i+1}-4\lambda\beta^r_{i}+(2\lambda-1)\beta^r_{i-1}
\right)\nonumber\\
&&\qquad+{n\over r_{max} }\left(
\lambda(\lambda+1)\beta^r_{i+1}+2(1-\lambda^2)\beta^r_{i}
+\lambda(\lambda-1)\beta^r_{i-1}\right) =0,
\end{eqnarray}
where
\begin{equation}\label{shiftbcouter2FD2}
\lambda\equiv {\tilde{r}_{max}-\tilde{r}_i \over \Delta \tilde{r}}.
\end{equation}
Here $\tilde{r}_{max}$ is the value of $\tilde{r}$ corresponding to
the point $r=r_{max}$, and Eqs.~(\ref{shiftbcouter2FD})
and~(\ref{shiftbcouter2FD2}) are evaluated at $i=i_{max}$.  Combining
Eqs.~(\ref{MinDistFD}) and~(\ref{shiftbcouter2FD}) to eliminate the
fictitious grid point $i_{max}+1$ yields a boundary condition for
$\beta^r_{i_{max}}$ in terms of $\beta^r_{i_{max}-1}$.

To impose the boundary condition~(\ref{shiftbcinner}) at the horizon,
we use the finite-difference expression
\begin{equation}\label{shiftbcinnerFD}
\lambda(\lambda+1)\beta^r_{i+1}+2(1-\lambda^2)\beta^r_{i}
+\lambda(\lambda-1)\beta^r_{i-1} 
=2\left.N\over A\right|_{\tilde{r}=\tilde{r}_{\rm AH}},
\end{equation}
where
\begin{equation}\label{shiftbcinnerFD2}
\lambda\equiv {\tilde{r}_{\rm AH}-\tilde{r}_i \over \Delta \tilde{r}}.
\end{equation}
Here $\tilde{r}_{\rm AH}$ is the $\tilde{r}$-coordinate location of
the apparent horizon, and Eqs.~(\ref{shiftbcinnerFD})
and~(\ref{shiftbcinnerFD2}) are evaluated at the value of $i$ such
that $i-1$ is the innermost grid point that lies outside the
horizon. Combining Eqs.~(\ref{MinDistFD}) and~(\ref{shiftbcinnerFD})
to eliminate the grid point at $i+1$ yields a boundary condition for
$\beta^r_{i-1}$ in terms of $\beta^r_i$. 

We currently use two methods to solve the linear system of equations
that results from Eqs.~(\ref{MinDistFD}), (\ref{shiftbcouter2FD}),
and~(\ref{shiftbcinnerFD}).  We use a standard tridiagonal
algorithm\cite{numrec_f} when running in serial, or when running in
parallel using a small number of processors.  Although very efficient,
this is a serial algorithm, so it begins to dominate the computation
time as the number of processors increases.  For a large number of
processors, we instead use a multigrid
technique\cite{numrec_f,Frontiers:Cook} which is parallelized with the
help of the DAGH system.

\section{Outer Boundary} \label{sec:outerbdry}

The only variables that require a boundary condition at the outer
boundary $r=r_{max}$ are the quantities $\lr$, $\lt$, $\mrr$, $\mrt$,
$\aor$, and $\arr$, which propagate along the ingoing and outgoing
null characteristics. All other variables evolve along the
characteristics parallel to the vector $\vec{\tilde{t}}$, and are
never differentiated with respect to $\tilde{r}$ in
Eqs.~(\ref{rewriteevolution}).

To apply a boundary condition on these variables, we assume that for
large $r$, the function $f$ for which we need a boundary condition
behaves like
\begin{equation} \label{oorbehavior}
f = {C_1\over r^n} + {C_2\over r^{n+1}}+\cdots
\end{equation}
for some integer $n$ and constants $C_j$.  We then impose
\begin{equation}\label{robinBC}
r{\partial^2\over \partial r^2}(r^nf) 
+2\dr(r^nf) = 0
\end{equation}
at the outer boundary.  This boundary condition truncates the
series~(\ref{oorbehavior}) after the second term, that is, it uses the
correct values of $C_1$ and $C_2$, but sets $C_j$ to zero for $j>2$.
We set the value of $n$ for each variable according to its analytic
falloff rate for the static solution.

We call this boundary condition an ``extended Robin'' method because
of its similarity to the familiar Robin boundary condition
\begin{equation}\label{realrobinBC}
\dr(r^nf) = C_0,
\end{equation}
which one often imposes on a function that behaves like
\begin{equation} \label{realrobinoorbehavior}
f = C_0 + {C_1\over r^n} + \cdots.
\end{equation}

Because Eq.~(\ref{robinBC}) involves approximating the large-$r$
behavior of the function $f$, it introduces errors that are no longer
second-order in the grid discretization, but fall off rapidly as one
increases the radius of the outer boundary.  These errors result
solely from the fact that Eq.~(\ref{robinBC}) is not exactly satisfied
by the initial data.  To obtain a second-order convergent scheme, we
effectively eliminate these approximation errors by increasing
$r_{max}$ until the second-order error terms dominate.

Because we use a cell-centered grid, the outer boundary is located at
half a grid spacing beyond the outermost grid point:
$r_{max}=r_{i_{max}+1/2}$.  Therefore, to ensure second-order
convergence in $\Delta r$, we must impose Eq.~(\ref{robinBC}) not at
the outermost grid point $r_{i_{max}}$, but at the outer boundary
$r_{max}$. This is necessary because $r_{i_{max}}$ depends on the grid
spacing $\Delta r$.

We impose Eq.~(\ref{robinBC}) by using the following finite-difference
approximations for the first and second derivatives:
\begin{mathletters}\label{robinBCFD}
\begin{eqnarray}
{\partial^2\over \partial r^2}Y_{i+1/2} &=&
{1\over 24 (\Delta r)^2}(7Y_{i-3}-40Y_{i-2}+102Y_{i-1}-112Y_{i}+43Y_{i+1})
\nonumber \\ &&
{}+{\mathcal O}(\Delta r)^3,\\
\dr Y_{i+1/2} &=&
{1\over 24 \Delta r}(Y_{i-2}-3Y_{i-1}-21Y_{i}+23Y_{i+1})
+{\mathcal O}(\Delta r)^3.
\end{eqnarray}
\end{mathletters}
Substituting these expressions in Eq.~(\ref{robinBC}) with $i=i_{max}$
yields an equation that can be solved for the quantity $r^n f$ at the
fictitious grid point $i=i_{max}+1$.  Using this quantity, we can now
evaluate the usual one-sided first derivative operator in
Eqs.~(\ref{macormackwavep1}--\ref{macormackwavep2}) at the outermost
grid point $i=i_{max}$.

This boundary condition needs to be applied only during the predictor
step of the Macormack scheme discussed in
section~\ref{sec:macormack}. This is because the corrector step uses a
backwards one-sided first derivative operator that is well-defined for
the outermost grid point.  Therefore, we need not worry about whether
we use $r$ or $\tilde{r}$ in Eqs.~(\ref{robinBC})
and~(\ref{robinBCFD}), because the normal and computational coordinate
systems coincide at the beginning of the predictor step.

The above boundary condition works well in the case of a spherically
symmetric, static solution, but it does not properly handle wave
propagation.  This includes not only physical gravitational waves,
which are absent in spherical symmetry, but also wavelike gauge modes,
which are present in our numerical solution.  While there exist many
methods for imposing wavelike outer boundary conditions, we find for
this one-dimensional problem that it suffices to simply move the outer
boundary far from the origin.

\section{Diagnostic tests} \label{sec:tests}

In this section we present results of rigorous convergence tests
performed with our code.

In all cases below, we evolve a Schwarzschild black hole in spherical
symmetry. We begin each simulation with time-symmetric initial data
corresponding to the $v=0$ plane of the Kruskal diagram. The initial
three-metric is written in isotropic coordinates:
\begin{equation} \label{ID3metric}
{}^{(3)}ds^2 = \left(1+{M\over 2 r}\right)^4
\left[dr^2+r^2(d\theta^2+\sin^2\theta\,d\phi^2)\right],
\end{equation}
where $M$ is the mass of the black hole, 
so that the initial values of
$A$,
$B$,
$\gamr$,
$\grt$,
$\kr$,
$\kt$,
$\mrr$,
and
$\mrt$
are given by
\begin{eqnarray} \label{InitMetric}
A &=& B =  \left(1+{M\over 2 r}\right)^2 \\
\gamr  &=&  -{M\over r^2}\left(1+{M\over 2 r}\right)^{-1} \\
\grt  &=&  2\left(1+{M\over 2 r}\right)\left(1-{M\over 2 r}\right) \\
\kr &=& 
\kt =
\mrr =
\mrt = 0.
\end{eqnarray}
With harmonic slicing, one is free to choose the initial value of the
lapse function. We choose
\begin{equation}
N=\left(1+{M\over r}\right)^{-1},
\end{equation}
so that the lapse agrees with the canonical Schwarzschild expression
at large $r$.
This choice of lapse yields
\begin{eqnarray} \label{initLapse}
\ar  &=&  {M\over r^2} \left(1+{M\over r} \right)^{-1} \\
\arr &=& -{2M\over r^3}\left(1+{M\over r} \right)^{-2}
                       \left(1+{M\over 2r}\right)^{-1}
		       \left(1+{M\over 2r}-{M^2\over 4r^2}\right)
                       \\
\at  &=&  {M\over r^3}\left(1+{M\over r} \right)^{-1}
                      \left(1+{M\over 2r}\right)^{-5}
                      \left(1-{M\over 2r}\right),
\end{eqnarray}
and the constraint
equations~(\ref{sphersymconstraints}--\ref{sphersymconstraints2}) give
us
\begin{eqnarray} \label{initConstraints}
\lr &=& -{M^2\over r^4}\left(1+{M\over 2r}\right)^{-2}
                       \left(1+{M\over r} \right)^{-2}
                       \left(3+{3M\over r} +{M^2\over 2r^2}\right)
                       \\
\lt &=&  {M^2\over r^4}\left(1+{M\over 2r}\right)^{-6}
                       \left(1+{M\over r} \right)^{-1}
                       \left(1+{M\over 4r}\right)
                       \\
\mtr &=& \aor = 0\label{InitConstraintsLast}.
\end{eqnarray}

We truncate our numerical grid just inside the event horizon at
$r=M/2$, and evolve only the exterior region.

\subsection{Self-Consistent Convergence} \label{sec:tests.self}

We have designed our code to be second-order accurate in
both space and time, so that if we evolve for a fixed time
$t$, our accumulated truncation error should be ${\mathcal O}(\Delta r)^2$,
where $\Delta r$ is our grid discretization.  Any deviation
from second-order convergent behavior would indicate a mistake in one
of our methods or a coding error.

To test whether our code is indeed second-order convergent, we evolve
the same initial data for the same time $t$ using different grid
discretizations, and we compare the results.  Let $X(t;\Delta r)$
be the value of some quantity $X$ at time $t$, computed with grid
discretization $\Delta r$.
For second-order convergent behavior,
\begin{equation}\label{Xerror}
X(t;\Delta r) = X(t)_{\rm true} + (\Delta r)^2X(t)_{\rm error} +\ldots,
\end{equation}
where $X(t)_{\rm true}$ is the true solution, and
$(\Delta r)^2X(t)_{\rm error}$ is the leading-order truncation error.
Hence
\begin{mathletters}\label{Xdifferror}
\begin{eqnarray}
        X(t;\Delta r/2) - X(t;\Delta r)           
                    &=& -{3\over 4}(\Delta r)^2X(t)_{\rm error} +\ldots,\\
4 \left[X(t;\Delta r/4) - X(t;\Delta r/2)\right] 
                    &=& -{3\over 4}(\Delta r)^2X(t)_{\rm error} +\ldots,\\
16\left[X(t;\Delta r/8) - X(t;\Delta r/4)\right] 
                    &=& -{3\over 4}(\Delta r)^2X(t)_{\rm error} +\ldots,
\end{eqnarray}
\end{mathletters}
and so on, so that each of the left-hand sides of
Eqs.~(\ref{Xdifferror}) are equal to leading order.

\begin{figure}[!htb]
\begin{center}
\begin{picture}(250,250)
\put(0,0){\epsfxsize=3.5in\epsffile{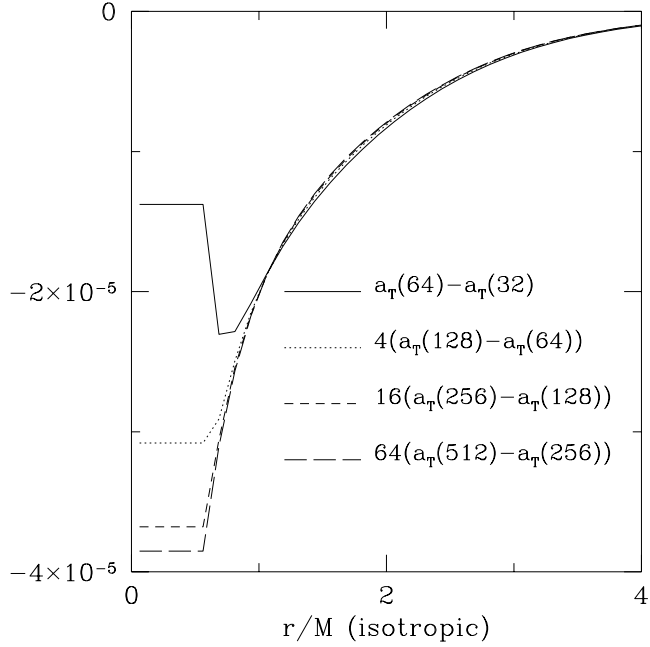}}
\end{picture}
\end{center}
\caption{Second-order error in the variable $\at$
at time $t=5.625M$.  The label $\at(N)$ denotes the value of $\at$
computed using $N$ grid points per unit $r$.  The apparent horizon is
located at approximately $r=M/2$.  All four plots coincide except at
small values of $r$, where higher-order error terms become
significant.  The portions of the graph at small $r$ where $\at$
appears to become constant represent grid points that are not included
in the evolution because they are inside
the apparent horizon.  The values of $\at$ at these grid points are
simply set to the value at the innermost significant point for
purposes of the plot.}
\label{fig:selfconvergeat}
\end{figure}

\begin{figure}[!htb]
\begin{center}
\begin{picture}(250,250)
\put(0,0){\epsfxsize=3.5in\epsffile{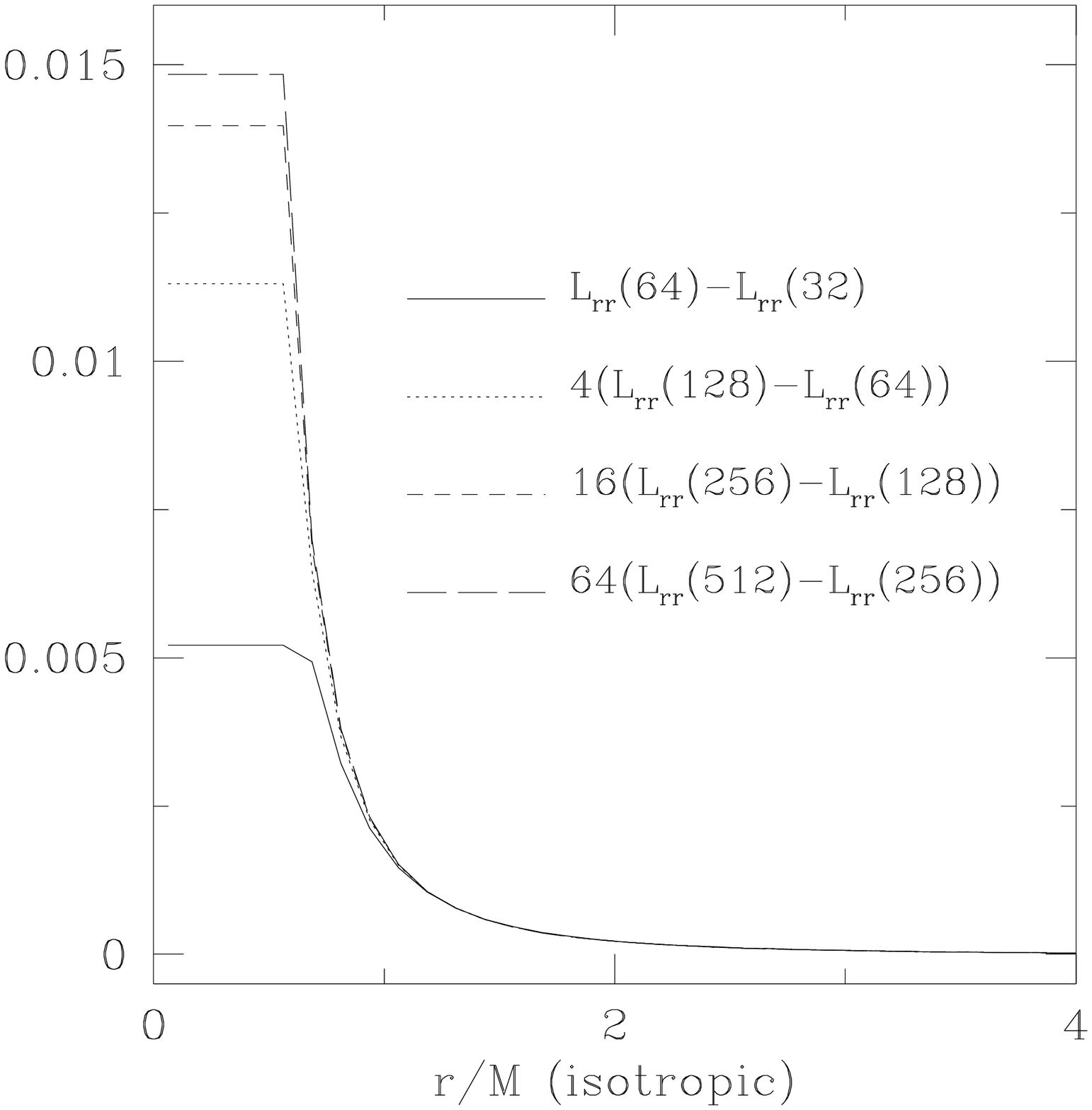}}
\end{picture}
\end{center}
\caption{Second-order error in the variable $\lr$ at time $t=5.625M$,
for the same case as shown in Figure~\ref{fig:selfconvergeat}.}
\label{fig:selfconvergelrr}
\end{figure}

Figures~\ref{fig:selfconvergeat} and~\ref{fig:selfconvergelrr}
demonstrate that the relation~(\ref{Xdifferror}) holds for our code.
These figures show results using five different grid resolutions, each
differing by a factor of $2$.  They were produced by evolving to
$t=5.625M$ using our AHBC method with a minimal distortion shift. This
time corresponds to $240$ time steps on the coarsest grid, and $3840$
time steps on the finest. The outer boundary is placed at $r=64M$.
Results from different resolutions are subtracted and scaled according
to Eq.~(\ref{Xdifferror}), and plotted together.

If only second-order error terms were present, the four plots in each
figure would coincide. This is indeed the case except at small $r$,
where the plots differ slightly because of third-order and higher
error terms, which are present but not explicitly given in
Eqs.~(\ref{Xdifferror}). These error terms vanish faster
than $(\Delta r)^2$ as $\Delta r\to 0$, as indicated by the
convergence of the plots toward each other as the grid resolution is
increased.

\begin{figure}[!htb]
\begin{center}
\begin{picture}(250,250)
\put(0,0){\epsfxsize=3.5in\epsffile{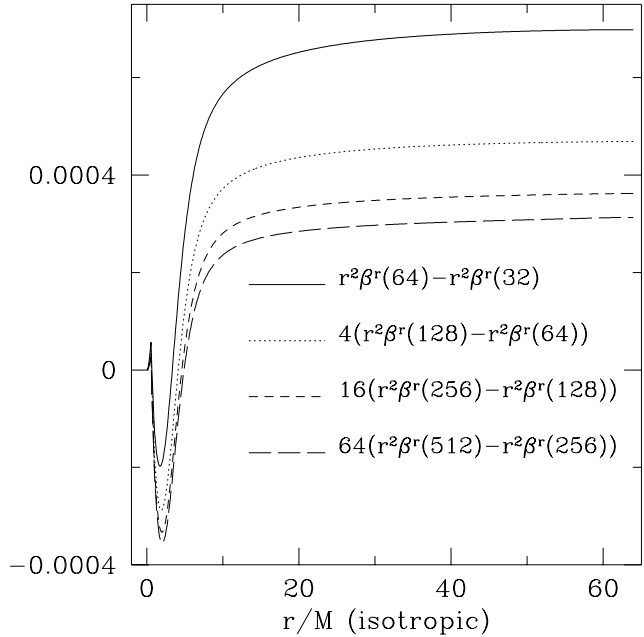}}
\end{picture}
\end{center}
\caption{Second-order error in the variable $r^2\beta^r$ at time
$t=5.625M$, for the same case as shown in
Figure~\ref{fig:selfconvergeat}.  }
\label{fig:selfconvergeshift}
\end{figure}

Figure~\ref{fig:selfconvergeshift} shows the second-order error in the
quantity $r^2\beta^r$, for the same case shown in
Figures~\ref{fig:selfconvergeat} and~\ref{fig:selfconvergelrr}.  This
demonstrates that our solution of the minimal distortion equation is
second-order convergent, and that it satisfies the outer boundary
condition to ${\mathcal O}(\Delta r)^2$. Higher-order
error terms are also present in this figure. They converge to
zero as the grid resolution is increased.

\subsection{Convergence of Constraint Equations}
\label{sec:tests.constraints}

Even if a numerical code is second-order convergent in the
self-consistent manner described above, it still may not converge to
the analytic solution.  To test whether this is the case, we evaluate
the left-hand sides of the constraint
equations~(\ref{sphersymconstraints}--\ref{sphersymconstraints2}),
which are not explicitly solved in the code except on the initial time
slice.  If we are indeed solving Einstein's equations, these
quantities should vanish like $(\Delta r)^2$.

\begin{figure}[!htb]
\begin{center}
\begin{picture}(250,250)
\put(0,0){\epsfxsize=3.5in\epsffile{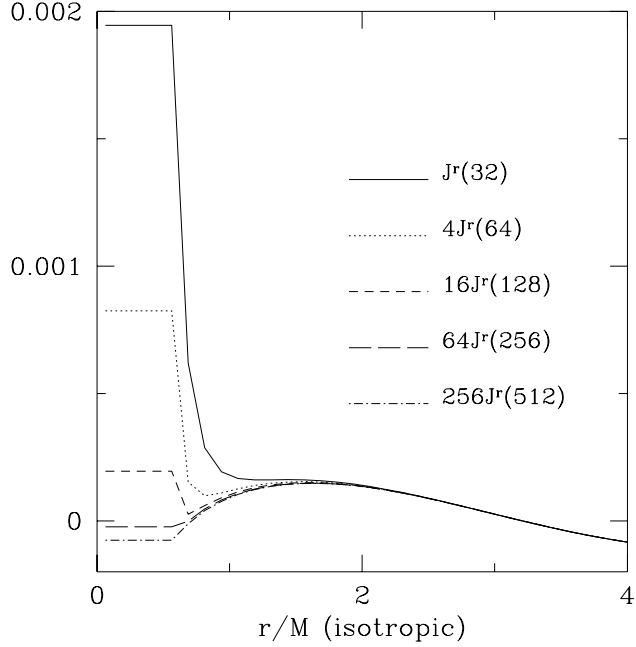}}
\end{picture}
\end{center}
\caption{Momentum constraint at time $t=5.625M$, for the same case as
shown in Figure~\ref{fig:selfconvergeat}.  In the plot labels,
$J^r(N)$ denotes the value of the left-hand side of
Eq.~(\ref{MomConst}) computed using $N$ grid points per unit $r$.}
\label{fig:momconstraint}
\end{figure}
 
Figure~\ref{fig:momconstraint} shows left-hand side of
Eq.~(\ref{MomConst}) for the same case shown in
Figures~\ref{fig:selfconvergeat}--\ref{fig:selfconvergeshift}, and for
five different grid resolutions.  For each finer grid discretization,
this quantity is multiplied by a factor of $4$, so that for strict
second-order convergence, the five plots should coincide. Because they
do coincide except for small $r$ where higher-order error terms are
significant, we see that the momentum constraint is satisfied to
${\mathcal O}(\Delta r)^2$.  The same is true for the other constraints.

There is an additional test we can use to determine whether we are
solving the correct equations: For a static black hole, the total mass
inside each radius $r$, which is a well-defined locally measurable
quantity in spherical symmetry, should be conserved.  The degree to
which mass conservation is violated provides a good indicator for the
overall accuracy of the code.  An invariant expression for the mass
inside a spherical surface of symmetry is \cite{problembook}
\begin{equation}\label{MassGeneral}
M\equiv\left({\mathcal A}\over 16\pi\right)
\left(1-{\nabla_a{\mathcal A}\nabla^a{\mathcal A}\over
16\pi{\mathcal A}}\right),
\end{equation}
where $\mathcal A$ is the area of the surface.
In terms of our variables, this expression reduces to
\begin{equation}\label{Mass}
M={Br\over 2}\left(1-\left(\grt\over 2A\right)^2+(Br\kt)^2\right).
\end{equation}
\begin{figure}[!htb]
\begin{center}
\begin{picture}(250,250)
\put(0,0){\epsfxsize=3.5in\epsffile{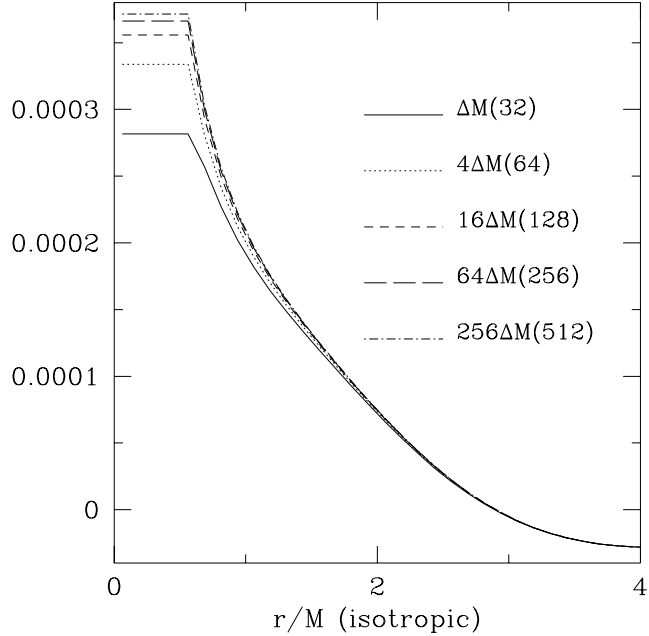}}
\end{picture}
\end{center}
\caption{The quantity $\Delta M$ at time $t=5.625M$, for the same case
as shown in Figure~\ref{fig:selfconvergeat}.}
\label{fig:mass}
\end{figure}
Figure~\ref{fig:mass} shows the quantity
\begin{equation}
\Delta M \equiv {M(t)-M(t=0)\over M(t=0)}
\end{equation}
at time $t=5.625M$ as calculated from Eq.~(\ref{Mass}), for five grid
discretizations. This quantity is multiplied by factors of $4$ for
each finer grid discretization.  The mass of the system is
conserved to ${\mathcal O}(\Delta r)^2$.

\section{Discussion} \label{sec:conclusion}

While the CBY formulation of Einstein's equations at first glance
looks more complicated than the usual ADM formulation, in most
respects it is actually much simpler, particularly from a numerical
point of view. Each equation in the CBY system is either a wave
equation with an advective term, or a simple advective equation.  In
the $\tilde{x}^i$ coordinate system that is used for causal
differencing, the equations are even simpler: They are either wave
equations or first-order ordinary differential equations in time.
Admittedly, the right-hand sides of these equations are complicated
and nonlinear, but because these right-hand sides contain no
derivatives, they do not make numerical solution of the equations any
more difficult.  The large number of evolution equations in the CBY
formalism is also not a serious drawback because the equations are all
of the same form, so they can all be solved by the same method.

Unlike a previous work\cite{scheel95a} in which we forced the horizon
to sit at a particular grid point, here we allow the horizon to lie at
some arbitrary location on the grid.  This is a closer approximation
to what we expect in the three-dimensional case, where one would most
likely use Cartesian coordinates to describe a spherical hole.
Likewise, in the same work we imposed explicit boundary conditions on
the apparent horizon in order to solve elliptic constraint equations
outside the hole. Here we do not impose explicit boundary conditions
(except for the shift via Eq.~(\ref{shiftbcinner})) because in the
three-dimensional case it is not only difficult to impose such
conditions on a nonspherical boundary, but also the number of boundary
conditions needed for solving all of the constraints in three
dimensions exceeds the number of boundary conditions available at the
horizon.

A key milestone in three-dimensional black hole simulations is the
ability to stably move a hole through the numerical grid. This is
arguably a necessary precursor to simulating binary orbits.
Techniques for moving holes through the grid could be tested in
spherical symmetry using our code, and we plan such tests in the
future. To preserve spherical symmetry, the motion in this case would
be the expansion or contraction (in coordinate space) of the hole
rather than the translation of the hole's center, but since the
location of the center is irrelevant for AHBC methods, many of the
same methods should apply to both cases.

The code as described here tends to lose accuracy at times greater
than 10 or 20 $M$, where $M$ is the mass of the black hole.  Because
this behavior is relatively independent of numerical details such as
grid resolution, we believe that it may be due to either gauge modes
or unphysical rapidly-growing solutions of the evolution equations
that do not satisfy the constraints.  One way of solving this problem
is to enforce several of the constraint equations after each time
step.  With this modification, we can integrate past $t=1000M$.
However, the details of constraint-violating solutions, gauge modes,
and methods for dealing with them are beyond the scope of this
paper and will be dealt with elsewhere\cite{scheel_etal97b}.

\acknowledgments
We thank Manish Parashar for his help with the DAGH system, and Andrew
Abrahams, Charles Evans, Ed Seidel, Wai-Mo Suen, and James York for
helpful discussions. This work was supported by the NSF Binary Black
Hole Grand Challenge Grant Nos. NSF PHY 93-18152/ASC 93-18152 (ARPA
supplemented), NSF Grant PHY 94-08378 at Cornell, and
NSF Grant AST 96-18524 and NASA Grant NAG 5-3420 at Illinois.

\end{document}